\documentclass[prl,twocolumn,showpacs,preprintnumbers,amsmath,amssymb,superscriptaddress]{revtex4-2}
\usepackage{graphicx}
\usepackage{bm}
\usepackage{dcolumn}
\usepackage{color}
\usepackage{bbm}
\usepackage{amsmath}
\usepackage{amssymb}
\usepackage{braket}
\usepackage[mathlines]{lineno}
\usepackage[colorlinks=true,citecolor=blue,linkcolor=blue,urlcolor=blue]{hyperref}
\begin{document}

\title{Quantum Phase Transitions in Optomechanical Systems}

\author{Bo \surname{Wang}}
\affiliation{School of Physics, Sun Yat-sen University, Guangzhou 510275, China}
\author{Franco \surname{Nori}}
\affiliation{Theoretical Quantum Physics Laboratory, Cluster for Pioneering Research, RIKEN, Wako-shi, Saitama 351-0198, Japan}
\affiliation{Center for Quantum Computing, RIKEN, Wako-shi, Saitama 351-0198, Japan}
\affiliation{Department of Physics, University of Michigan, Ann Arbor, Michigan 48109-1040, USA}
\author{Ze-Liang \surname{Xiang}}
\altaffiliation[xiangzliang@mail.sysu.edu.cn]{}
\affiliation{School of Physics, Sun Yat-sen University, Guangzhou 510275, China}

\begin{abstract}
In this letter, we investigate the ground state properties of an optomechanical system consisting of a coupled cavity and mechanical modes. An exact solution is given when the ratio $\eta$ between the cavity and mechanical frequencies tends to infinity. This solution reveals a coherent photon occupation in the ground state by breaking continuous or discrete symmetries, exhibiting an equilibrium quantum phase transition (QPT). In the $U(1)$-broken phase, an unstable Goldstone mode can be excited. In the model featuring $Z_2$ symmetry, we discover the mutually (in the finite $\eta$) or unidirectionally (in $\eta \rightarrow \infty$) dependent relation between the squeezed vacuum of the cavity and mechanical modes. In particular, when the cavity is driven by a squeezed field along the required squeezing parameter, it enables modifying the region of $Z_2$-broken phase and significantly reducing the coupling strength to reach QPTs. Furthermore, by coupling atoms to the cavity mode, the hybrid system can undergo a QPT at a hybrid critical point, which is cooperatively determined by the optomechanical and light-atom systems. These results suggest that this optomechanical system complements other phase transition models for exploring novel critical phenomena.

\end{abstract}

\maketitle

\emph{Introduction.---}Quantum phase transitions have garnered significant attention for their crucial role in comprehending the various intricacies involved in the evolution of matter phases~\cite{sachdevQPT,1997SondhiContinuous,vojtaQPT}. Equilibrium QPTs can be characterized by the closing spectral gap and the emergence of degenerate ground states due to the spontaneous breaking of a continuous (or discrete) symmetry. For systems possessing continuous $U(1)$ symmetry, such as the Bose-Hubbard model, their phase transitions~\cite{2012Endres,2023wangProbing} are attracting intense interest due to their involvement in exhibiting notable physics of the Anderson-Higgs mechanism~\cite{1963AndersonPlasmons,1964HiggsBroken}.

In quantum optics, the well-known Dicke model, describing the coupling between a cavity mode and $N$ two-level systems, can exhibit a superradiant QPT in the thermodynamic limit, $N\rightarrow \infty$, where the bosonic mode gains the occupation of macroscopic coherence in the ground state~\cite{1973HeppAndLieb,1973WangAndHioe,2003EmaryAndBrandes}. Recently, this QPT was also predicted in the Rabi model in the classical oscillator limit~\cite{2013Ashhab,2015HwangQPT}, where the ratio between frequencies of the atomic transition and the cavity mode approaches infinity. These two fundamental spin-boson models can be reduced to a model with a $U(1)$ symmetry by dropping the so-called counter-rotating terms, resulting in the emergence of Goldstone modes~\cite{2016HwangQuantum,2014BaksicControlling}. In the ultrastrong-coupling regime, these terms usually cannot be ignored~\cite{2019frisk,2019Forn}; however, recent progress indicated that this requirement could be realized by engineering the light-matter interaction in circuit-QED systems~\cite{2014BaksicControlling} or through quantum simulation techniques~\cite{2017Langford,2014Mezzacapo,2014Georgescu}.

Previous investigations of the superradiant QPT have predominantly focused on the spin-boson model, with either infinite~\cite{2007DimerProposed,2010baumannDicke,2011BaumannExploring,2010NagyDicke,2015klinderdynamical,2006vidalfinite,2012BakemeierQuantum,2004LambertEntanglement,2020GuerciSuperradiant,2019MazzaSuperradiant,2022ChiriacCritical,2022ZhaoFrustrated} or finite components~\cite{2017PueblaProbing,2017LiuUniversal,2020GarbeCritical,2020ZhuFinite,2021ChenExperimental,2021CaiObservation,2021ZhangQuantumPhase,2022FallasUnderstanding}. This naturally leads to an intriguing question of whether the boson-boson model can similarly exhibit a rich range of quantum phases, thereby improving our understanding of QPTs. The optomechanical system considered here is a boson-boson model, describing the interaction between a cavity mode and a mechanical oscillator through radiation pressure~\cite{2014AspelmeyerCavity,2022BarzanjehOptomechanics}. With recent developments of strong single-photon optomechanical couplings~\cite{2014HeikkilEnhancing,2015ViaStrong,2015LuSqueezed,2017ShevchukStrong,2018NeumeierReaching,2020KounalakisFlux,2022ManninenEnhancement,2015CavityPirkkalainen,2022CarlonEnhanced,2019RodriguesCoupling}, it enables the investigation of quantum nonlinear effects in optomechanical systems, such as preparing non-classical states~\cite{1997BosePreparation,2011NunnenkampSingle,2011RablPhoton,2013LiaoPhoton,2012QianQuantum,2016RiedingerNon} and observing the dynamical Casimir effect~\cite{2018MacriNonperturbative,2023WangCoherent,2019StefanoInteraction,2009JohanssonDynamical,2012NationColloquium}. These nonlinear effects lead to energy-level repulsion and attraction that can cause the degeneration of the lowest levels, implying the emergence of QPTs.

In this letter, we study the ground state properties of optomechanical systems based on models with either $U(1)$ symmetry or $Z_2$ symmetry and give the analytical solutions of their equilibrium QPTs. The exact solution for a model possessing $U(1)$ symmetry reveals the instability of the ground state and indicates the emergence of a Goldstone mode. For the model with $Z_2$-symmetry, in the finite $\eta$, displacing the phonon space can generate a pair of mutually dependent squeezed vacuum between the cavity and mechanical modes via radiation pressure; however, in the limit, $\eta \rightarrow \infty$, such dependence is unidirectional. Interestingly, applying a squeezed field to drive the cavity with the required squeezed parameter, we find that the features of QPTs can be remarkably influenced: the region where the $Z_2$-broken phase occurs is alterable, and the coupling strength to reach the critical point can be significantly reduced. In addition, interacting with atoms, the hybrid system will have a hybrid critical point, where the optomechanical and light-atom components of the system could cooperatively determine the critical phenomena.


\emph{Model.---}We consider a typical optomechanical system consisting of a cavity with a movable mirror. The system Hamiltonian~\cite{1995LawInteraction} can be written as 
\begin{equation}\label{eq9}
	H=\omega_c a^{\dag}a+\omega_m b^{\dag}b+g(a+a^{\dag})^{2}(b+b^{\dag}).
\end{equation}
where the annihilation operator $a(b)$ denotes the optical (mechanical) mode with the resonate frequency $\omega_c (\omega_m)$, and $g$ is the strength of single-photon optomechanical coupling. For most experiments to date~\cite{2014AspelmeyerCavity}, the Hamiltonian $H_{\text{om}}=\omega_c a^{\dag}a+\omega_m b^{\dag}b+2g a^{\dag}a(b+b^{\dag})$ is a sufficient good approximation. While the single-photon coupling strength is increasing, the terms $g(a^2+a^{\dag 2})(b+b^{\dag})$, describing the creation and annihilation of photon pairs~\cite{2018MacriNonperturbative,2023WangCoherent}, become considerable. Obviously, $2g a^{\dag}a(b+b^{\dag})$ and $g(a^2+a^{\dag 2})(b+b^{\dag})$ have distinct effects on the photon occupation in the ground state due to the different symmetries of the cavity mode.


\emph{Photon occupations in ground state.---}
First we focus on the simplest model described by the Hamiltonian $H_{\text{om}}$. Let us denote $|n\rangle$ as an $n-$photon Fock state and $|k\rangle$ as an $k-$phonon Fock state, respectively. By performing a unitary transformation with $U=\exp[-(g/\omega_m)a^{\dag}a(b^{\dag}-b)]$, the Hamiltonian becomes $\bar{H}_{\text{om}}=(U^{\dag}H_{\text{om}}U)/\omega_c= a^{\dag}a+\eta^{-1}b^{\dag}b-(1/\kappa^2)a^{\dag}aa^{\dag}a$ with a dimensionless coupling strength $\kappa=\sqrt{\omega_c \omega_m}/2g$ and a frequency ratio $\eta=\omega_c /\omega_m$, where the eigenstate and the eigenvalue can be described by the ket $|n,k\rangle$ and $E_{\text{om}}=n-(n^2/\kappa^2)+\eta^{-1} k$, respectively. In the displaced basis, $\bar{H}_{\text{om}}$ has a conserved operator, $\bar{P}=\exp[i\theta N]$ with $\theta \in [0,2\pi]$ and $N=a^{\dag}a+b^{\dag}b$, such that $[\bar{H}_{\text{om}},\bar{P}]=0$, satisfying a $U(1)-$continuous symmetry. Going back to the original basis, the corresponding conserved operator and eigenstates can be written as
\begin{equation}\label{eq3}
P=U\bar{P}U^{\dag}=e^{{i\theta(a^{\dag}a+b^{\dag}b+\frac{2g}{\omega_m}a^{\dag}a(b+b^{\dag})+\frac{4g^2}{\omega_m}a^{\dag}aa^{\dag}a)}}
\end{equation}
and $|\psi \rangle=|n,k_n\rangle=D(n\sqrt{\eta}/ \kappa) |n,k\rangle$, with $D(x)=\exp[x(b-b^{\dag})]$, respectively. 



\begin{figure}[tpb]
\centering
\includegraphics[width = 0.98  \linewidth]{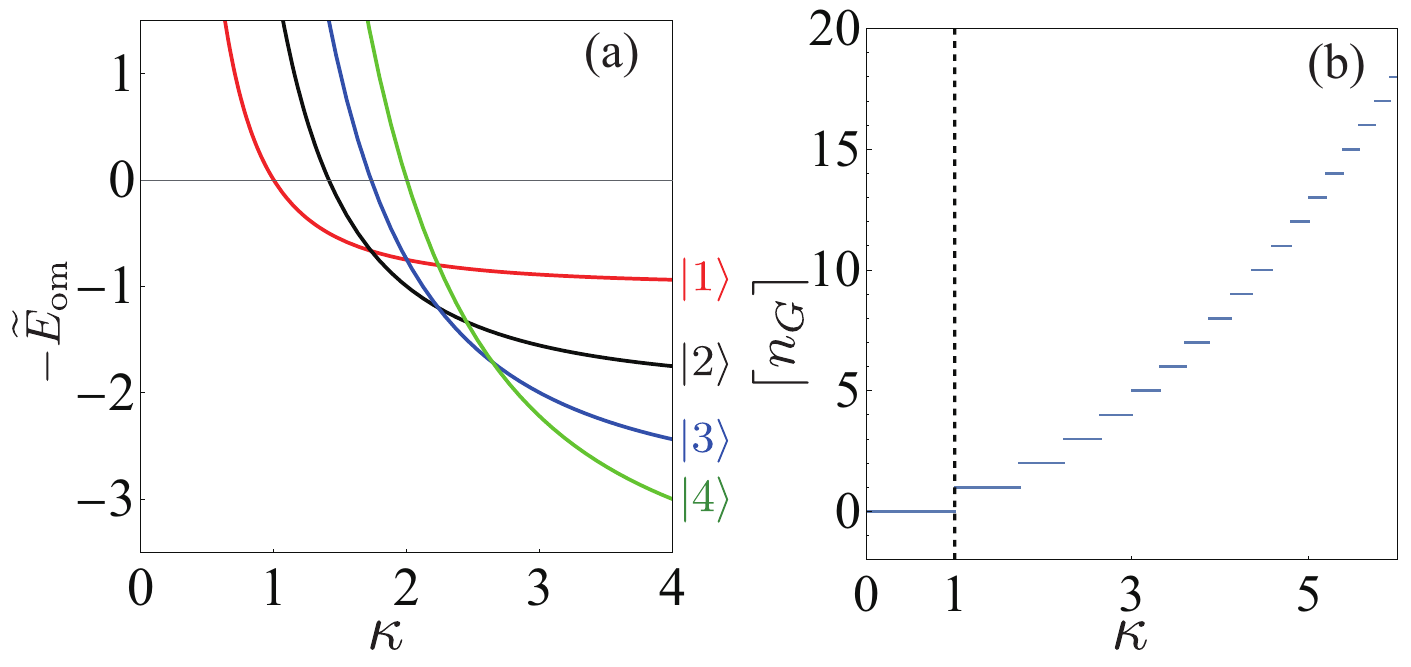}
\caption{Analytic solution of the Hamiltonian $H_{\text{om}}$. (a) Level crossings of the low-energy state near the ground state for a frequency ratio $\eta \rightarrow \infty$. 
(b) The total number of excitations in the ground state for a frequency ratio $\eta \rightarrow \infty$. 
} 
\label{V_om1}	 	
\end{figure}

\begin{figure}[tpb]
\centering
\includegraphics[width = 0.98  \linewidth]{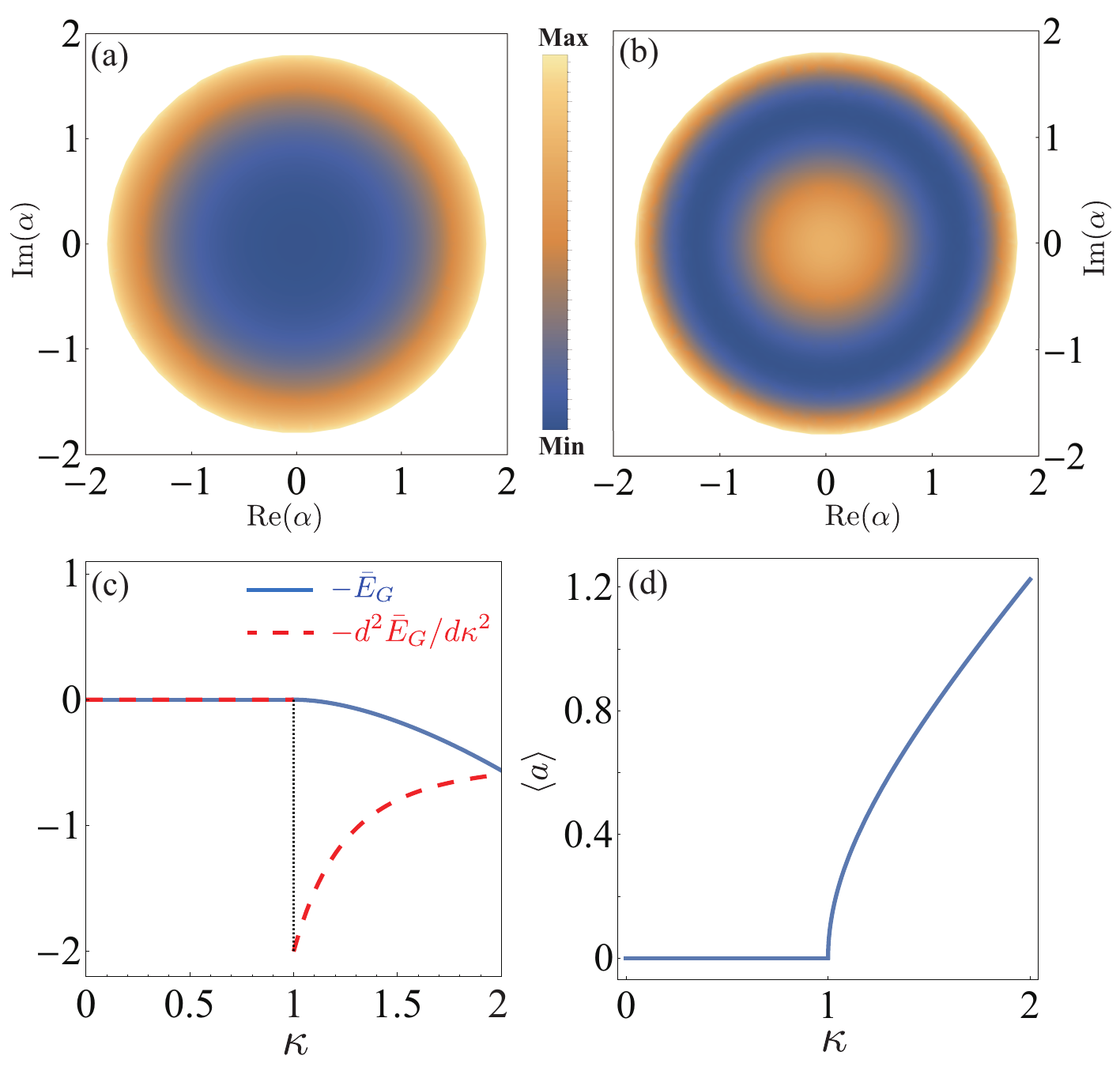}
\caption{Mean-field solution of the Hamiltonian $\bar{H}_{\text{om}}$. Mean-field energy of the ground state when $\kappa=0.5$ (a) and when $\kappa=2$ (b). (c) Ground state energy $\bar{E}_G$ (solid blue line) and its second derivative $d^2 \bar{E}_G/d \kappa^2$ (red dashed line). (d) Coherence $\langle \alpha \rangle$ of the cavity field in its ground state. 
}
\label{V_om2}	 	
\end{figure}

Now we consider a particular limit, i.e., $\eta \rightarrow \infty$. The Hamiltonian can be rewritten as 
\begin{equation}\label{eq5}
\widetilde{H}_{\text{om}}=a^{\dag}a-\frac{1}{\kappa^2}a^{\dag}aa^{\dag}a
\end{equation}
with eigenvalues $\widetilde{E}_{\text{om}}=n[1-(n/\kappa^2)]$, exhibiting an anharmonic spectrum. To obtain a well-defined ground state in Eq.~\eqref{eq5}, $a^{\dag}a$ needs to be treated as a perturbation, which means the system energy is dominated by the negative nonlinear term with $(-1/\kappa^2)n^2$~\cite{SM}. Therefore, we can perform the transformation $\widetilde{E}_{\text{om}} \rightarrow -\widetilde{E}_{\text{om}}$ shown in Fig.~\ref{V_om1} (or $\bar{E}_{G} \rightarrow -\bar{E}_{G}$ shown in Fig.~\ref{V_om2}), which only changes the reference frame but allows us to conveniently capture the nature of the ground state.

For $\kappa<1$, the well-defined ground state of $\widetilde{H}_{\text{om}}$ is the vacuum state $|0\rangle$, until at $\kappa=1$ there occurs a level crossing between $|0\rangle$ and $|1\rangle$. After that, the ground state meets a series of level crossings between the states $|n\rangle$ and $|n+1\rangle$, as shown in Fig.~\ref{V_om1}(a). From $\widetilde{E}_{\text{om}}^{(n+1)}-\widetilde{E}_{\text{om}}^{(n)}=0$, the number of photons occupying the ground state, denoted as $n_G$, can be described by 
\begin{equation}\label{eq6}
	n_G=\frac{\kappa^2-1}{2}.
\end{equation}
Obviously, as $\kappa$ increases, the quantity $\lceil n_G \rceil$, which counts the number of photons using the ceiling function $\lceil ~ \rceil$, experiences a discrete step-wise ascent, as shown in Fig.~\ref{V_om1}(b), showing the trend of photon occupations in the ground state.



\emph{Results in mean-field approach.---}Next we use the mean-field approach, a semiclassical approximation, to observe the distribution of the ground-state energy. By displacing both modes $a$ and $b$ in $\bar{H}_{\text{om}}$, with respect to their corresponding mean values $\alpha$ and $\beta$, and neglecting the fluctuations, we can obtain the mean value of the ground state energy in terms of $\alpha$ and $\beta$, 
\begin{equation}
\bar{E}_G=|\alpha|^2+\eta^{-1}|\beta|^2-\frac{1}{\kappa^2}(|\alpha|^2+|\alpha|^4).
\end{equation}
When $\kappa<1$, the ground state energy has a single minimum at $\alpha=\beta=0$, where the energy is $\bar{E}_G=0$, as shown in Fig.~\ref{V_om2}(a). For $\kappa>1$, the distribution of the ground-state energy abruptly changes as a profile of Mexican hat shown in Fig.~\ref{V_om2}(b), where the ground state in $\alpha=0$ becomes unstable and will fall into the stable one with the energy minima occurring at 
\begin{equation}\label{eq7}
\alpha=\pm e^{i\theta}\sqrt{\frac{1}{2}(\kappa^{2}-1)}~ \text{and}~
\beta=0.
\end{equation}
Because the phase $\theta$ can be an arbitrary value from 0 to $2\pi$, the ground states are infinitely degenerate with the energy minima lying on a circle, which is known as Goldstone mode~\cite{1962GoldstoneBroken} with broken $U(1)$-symmetry.

For $\kappa>1$, the ground state energy becomes $\bar{E}_G(\kappa)=(1/4)(\kappa^2+\kappa^{-2}-2)$. When the ground state energy is continuous, there is a discontinuity in the second derivative of $\bar{E}_G(\kappa)$ at $\kappa=1$, revealing a second-order phase transition [Fig.~\ref{V_om2}(c)]. In Eq.~\eqref{eq7}, the nonzero values of $\alpha$ mean that the ground state in $\kappa >1$ has a nonzero coherence of the cavity mode, which is an order parameter of the phase transition [Fig.~\ref{V_om2}(d)] and indicates the spontaneous breaking of $U(1)$ symmetry. Based on nonzero coherence, we can obtain the photon occupation in the ground state, namely, $\langle a^{\dag}a \rangle_{G}=(\kappa^2-1)/2$, which is consistent with the solution of Eq.~\eqref{eq6}. 

Although the results in Fig.~\ref{V_om2} can significantly indicate that the model $H_{\text{om}}$ can undergo a phase transition through the critical point $\kappa=1$, the lack of thermodynamic limit in Eq.~\eqref{eq7} implies that spontaneous symmetry breaking would be restored by quantum fluctuations. Nevertheless, for a finite-component system~\cite{2003EmaryAndBrandes} or finite-frequency system~\cite{2015HwangQPT}, quantum fluctuations still require a finite time to restore the symmetry of the real ground state~\cite{wenQFT}. Therefore, it is possible to observe the above results, even though the symmetry-breaking ground state is not stable.


\emph{Squeezed vacuum in photon and phonon modes.---}When the Hamiltonian includes $g(a^2+a^{\dag 2})(b+b^{\dag})$, it is difficult to determine the symmetry of the system. However, we find that the Hamiltonian $H$ commutes with the operator $P_{\text{c}}=\exp(i\pi a^{\dag}a)$, indicating that the cavity mode possesses an independent $Z_2$-parity symmetry. Accordingly, one may execute a displacement transformation on the mechanical mode with the unitary operator, $U_1=\exp[-(g/\omega_m)(b^{\dag}-b)]$, and the transformed Hamiltonian $U_1^{\dag}HU_1$ becomes
\begin{equation}\label{eq10}
\begin{split}
\bar{H}= \,& \, \omega_c a^{\dag}a+\omega_m b^{\dag}b-\frac{2g^2}{\omega_m}(a+a^{\dag})^2+\frac{g^2}{\omega_m } \\
\,& \, +g (a^2+a^{\dag 2})(b+b^{\dag})+2g a^{\dag}a(b+b^{\dag}).
\end{split}
\end{equation}
Note that this displacement elicits an anti-squeezing term, the third one in Eq.~\eqref{eq10}, which could induce non-analytic behavior of the cavity mode at one point upon changing the coupling strength~\cite{SM}. Such behavior will give feedback to the mechanical oscillator by radiation pressure and potentially influence the phonon mode's behavior. To gain insight into the underlying physics, one can perform a transformation on the Hamiltonian $\bar{H}$~\cite{SM} with the unitary operator, $U_2=\exp[-(g/2\omega_c)(b^{\dag}+b)(a^{\dag 2}-a^2)]$. This transformed Hamiltonian, $\widetilde{H}=U_2^{\dag}\bar{H}U_2$, although not yet able to be diagonalized, may be analyzed via a variational method.

We propose a trial wave function $|\psi(r,s)\rangle=\mathcal{S}_a(r)\mathcal{S}_b(s)|0_a,0_b\rangle$ for $\widetilde{H}$, where $\mathcal{S}_y(x)=\exp[(x^2/2)(y^{\dag 2}-y^2)]$ is the squeezing operator with a bosonic operator $y\in[a,b]$ and a variational parameter $x \in [r,s]$ providing the energy function $\tilde{E}(r,s)$~\cite{SM}. Without loss of generality, we here require that the energy function is up to fourth order in $\gamma$ in the following analysis, where $\gamma=2\sqrt{2}g/\sqrt{\omega_c \omega_m}$ is a dimensionless coupling strength. By minimizing the energy with respect to $r$ and $s$, we can obtain
\begin{equation}\label{eq12}
e^{4r}=\frac{1+\frac{1}{4}\gamma^2 \eta^{-1}e^{2s}+\frac{1}{32}\gamma^4\eta^{-2}e^{4s}}{1-\gamma^2(1+\frac{3}{4}\eta^{-1}e^{2s}+\frac{1}{4}\gamma^2 \eta^{-1}e^{2s}+\frac{7}{32}\gamma^2 \eta^{-2} e^{4s})} 
\end{equation}
and
\begin{equation}\label{eq13}
e^{4s}=\frac{[\frac{3}{32}\gamma^4\eta^{-1}(8\sinh^2{r}+4)+\frac{1}{2}\gamma^4\eta^{-1}\sinh{2r}]e^{6s}+1}{[1-\gamma^2(\sinh^2{r}+\sinh{2r}+\frac{1}{2}+\frac{1}{4}\gamma^2 e^{2r})]}.
\end{equation} 

When $\eta$ is finite, nonzero and correlated solutions for the squeezing parameters $r$ and $s$ always exist if $\gamma<1$. This reflects the inseparable relation between the squeezed vacuum states of the cavity and mechanical modes induced by the radiation pressure. 

When $\eta \rightarrow \infty$, Eqs.~(\ref{eq12},\ref{eq13}) can be rewritten as $e^{4r}=1/(1-\gamma^2)$ and $e^{4s}=1/[1-\gamma^2(\sinh^2{r}+\sinh{2r}+\frac{1}{2}+\frac{1}{4}\gamma^2 e^{2r})]$, respectively. In this case, $r$ no longer relies on $s$, but $s$ still depends on $r$, indicating that the cavity mode carries an irreversible feedback to the mechanical mode. This arises from the deterministic symmetry of the cavity mode itself, independent of the mechanical mode, namely, $[H, P_{\text{c}}]=0$. Moreover, as $\gamma \rightarrow \gamma_c=1$, a transition arises, signifying a profound divergence in both $r$ and $s$. These divergences in the ground state imply the phase transitions of the system~\cite{2015HwangQPT,2003EmaryAndBrandes}, where the two-fold degeneracy occurs. Interestingly, the phase transition of the mechanical mode would suggest a hidden symmetry within $H$, which is cooperatively determined by the cavity and mechanical modes and would be broken in the ground state. 




\emph{Superradiant phase.---}To address the superradiant phase in the Hamiltonian $H$, it is necessary to recognize that $H_I=g(a+a^{\dag})^2(b+b^{\dag})$ is unbounded, which is an obstacle in locating the position of macroscopic coherence of the bosonic mode in the ground state~\cite{2020FelicettiUniversal}. One feasible way is to introduce a nonlinear quartic term to each bosonic mode, thereby obtaining a Hamiltonian
\begin{equation}\label{eq14}
\begin{split}
H_{\text{op}} =\,&\, N\omega_c a^{\dag} a+\omega_m b^{\dag}b+g(a^2+a^{\dag 2}+2a^{\dag}a)(b+b^{\dag})\\
 \,&\, +Ng(b+b^{\dag})+\frac{\epsilon_1}{N^2}a^{\dag}a^{\dag}aa+\frac{\epsilon_2}{N^2}b^{\dag}b^{\dag}bb,
\end{split}
\end{equation}
where $N$ is a macroscopic factor. Given the factor $N$, the detuning between the two oscillators is determined. As $N$ becomes large, the resulting Hamiltonian in Eq.~\eqref{eq14} is equivalent to the one in Eq.~\eqref{eq9}, faithfully illustrating that the two nonlinear quartic terms are vanishingly small.

Next, we adopt a mean-field approach to examine the ground state energy of the Hamiltonian $H_{\text{op}}$ by displacing both modes $a$ and $b$ from their mean values, denoted as $\alpha$ and $\beta$. The expression of the ground state energy is given by
\begin{equation}\label{eq14energy}
E_G=N\omega_c \alpha^2+\omega_m \beta^2+2g\beta(N+4\alpha^2)+\frac{\epsilon_1}{N^2}\alpha^4+\frac{\epsilon_2}{N^2}\beta^4.
\end{equation}
To minimize the energy $E_G$, we let $\beta/\alpha^2 \sim 1$ and $\alpha \sim \sqrt{N}$, which yields 
\begin{equation}\label{eq14phase}
\alpha^2=\frac{N}{4}(\gamma^2-1)~\text{and}~\beta=-\frac{N\omega_c}{8g},
\end{equation}
where $\epsilon_1=\epsilon_2=(4\omega_m^2/\omega_c)(\gamma^6-\gamma^2)$. For $\gamma >1$, the nonzero coherence of the mode $\langle a\rangle=\pm \alpha$ indicates the superradiant phase with a spontaneously broken-parity symmetry $P_{\text{c}}$.


\emph{Modification to features of QPT.---}We will now show that by employing a controllable squeezed field of the cavity mode to drive the optomechanical system, it is possible to modify the characteristics of the QPT. To illustrate this point, we apply the transformation with the unitary operator, $S_{\zeta}=\exp{[(1/2)(\zeta^{*}a^2-\zeta a^{\dag 2})]}$, where $\zeta=\xi \exp(i\theta)$ represents the squeezed parameter, to the system Hamiltonian $H$ with a squeezing-driven term $\xi(a^{\dag 2}e^{-i\theta}+a^2e^{i\theta})$. The transformed Hamiltonian depends on the squeezing direction $\theta$ and the squeezing amplitude $\xi$~\cite{SM}, providing a channel to modify the ground state properties of the cavity mode.


First, we consider the squeezing direction $\theta=0$, where the Hamiltonian $H_{(\theta=0,\xi)}$ exhibits a similar structure to $H$. By performing a displacement transformation with $\widetilde{U}=\exp[-(g e^{-2\xi}/\omega_m)(b^{\dag}-b)]$, and taking the limit $\eta \rightarrow \infty$, the Hamiltonian $H_{(\theta=0,\xi)}$ can be diagonalized as $\widetilde{H}_{(\theta=0,\xi)}=2\varepsilon_{\xi} d^{\dag}d+\widetilde{E}_{\text{G}(\xi)}$, with $\varepsilon_{\xi}=\sqrt{(1/4)(1-\gamma^2 e^{-2\xi})}$. For $\xi =2\ln(\gamma)$, $\varepsilon_{\xi}$ simplifies to 
\begin{equation}\label{eq13.6}
	\varepsilon_{\xi \rightarrow 2\ln(\gamma)}=\sqrt{(1/4)(1-\gamma^{-2})},
\end{equation}
which is real only for $\gamma \geq 1$ and vanishes at $\gamma=1$. However, when $\xi=0$, i.e., without the squeezing-driven field, the eigenvalue is given by $\varepsilon_{\xi \rightarrow 0}=\sqrt{(1/4)(1-\gamma^2)}$, indicating the occurrence of the phase for $\gamma <1$. This result demonstrates that the region where the quantum phase occurs can be altered by driving the system with a squeezed field of the squeezing direction $\theta=0$ and an appropriate squeezing amplitude. This feature is remarkably unusual within the field of phase transitions, indicating the modifying capability on phase diagrams.

Furthermore, for the squeezing direction $\theta=\pi$, the corresponding eigenvalue is given by $\varepsilon_{\xi}=\sqrt{(1/4)(1-\gamma^2 e^{2\xi})}$, indicating that the optomechanical coupling strength required to reach the critical point can be exponentially reduced by increasing the squeezing amplitude $\xi$. This result suggests the possibility of the phase transition in the optomechanical system even with a common coupling strength. Of note, such a way cannot apply to the Rabi and Dicke models, as these models lack the inherent symmetry of the cavity mode itself.

\begin{figure}[tpb]
\centering
\includegraphics[width = 0.98  \linewidth]{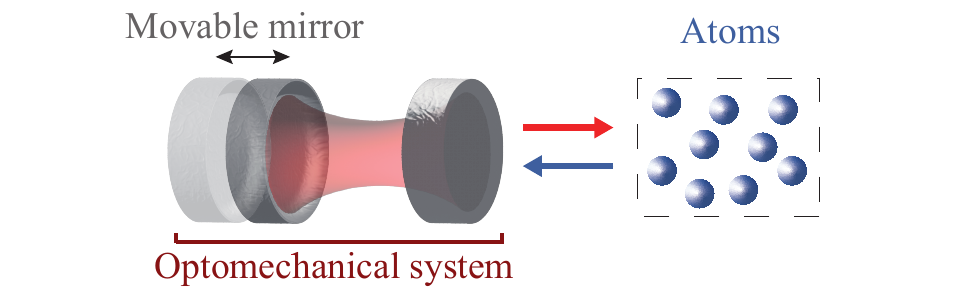}
\caption{Schematic of a hybrid system consisting of a cavity coupled to both a movable mirror and an atomic ensemble.}
\label{hybridSystemSchematic}	 	
\end{figure}

\emph{QPTs in hybrid systems.---}We now turn to an optomechanical system interacting with atoms in the cavity, as shown in Fig.~\ref{hybridSystemSchematic}. The results reveal that QPT can emerge in such a system, which has a \textit{hybrid} critical point cooperatively determined by the optomechanical system and the light-atom system. Here, we employ the Dicke model to describe the light-atom interaction. The Hamiltonian of this hybrid system~\cite{annotation2} can be written as 
\begin{equation}\label{hybrid-1}
H_{\text{h}} =H +\omega_a J_z+\frac{\lambda}{\sqrt{N_a}}(a+a^{\dag})(J_{+}+J_{-})+\chi(a+a^{\dag})^2
\end{equation}
with the angular momentum operator $J_z=(1/2)\sum_{i=1}^{N_a}\sigma_z^{(i)}$ and $J_{\pm}=\sum_{i=1}^{N_a} \sigma_{\pm}^{(i)}$, satisfying the commutation relation $[J_{-},J_{+}]=-2J_z$. The last term in Eq.~\eqref{hybrid-1} is so-called $\bm{A}^2$ term with $\chi=\alpha \lambda^2/\omega_a$~\cite{annotation1}. The coefficient $\alpha\geqslant 1$ ensures that the Dicke model has a precisely gauge-invariant Hamiltonian satisfying the TRK sum rule~\cite{1975RzaPhase,2020savastathomas}, while $\alpha=0$ corresponds to the standard Dicke model~\cite{2003EmaryAndBrandes}. After diagonalizing the Hamiltonian $H_{\text{h}}$~\cite{SM}, the spectrum shows that the excitation energy of the lowest branch $\epsilon_{-}$ vanishes at 
\begin{equation}
\mu^2(1-\alpha)+\gamma^2=1,
\end{equation}
indicating a quantum critical point of the hybrid system, where $\mu=2\lambda/\sqrt{\omega_a \omega_c}$ is a dimensionless coupling strength of the light-atoms interacting system.

When neglecting the $\bm A^2$ term, i.e., $\alpha=0$, the critical point can be expressed as $\mu^2+\gamma^2=1$. This formula contains two distinct types of dimensionless coupling strength, $\mu$ and $\gamma$, which correspond to the critical points in the standard Dicke model ($\mu^2=1$) and the optomechanical model ($\gamma^2=1$), respectively. Thus, the hybrid quantum system features a \textit{hybrid} critical point that separates the normal and superradiant phases, where the critical phenomena are dominated by both the light-atom and optomechanical systems.

When $\alpha=1$, the \textit{hybrid} critical point becomes $\gamma^2=1$. In this scenario, one trend of the closing spectrum gap, induced by the light-atom interaction, is completely suppressed by the $\bm A^2$ term, in agreement with the no-go theorem~\cite{1975RzaPhase}. However, the other trend, caused by the interaction between the cavity mode and the mechanical oscillator, results in a gapless spectrum, where the critical point only depends on $\gamma$. Nevertheless, the energy spectrum for the superradiant phase remains influenced by $\mu$~\cite{SM}.




\emph{Conclusion and outlook.---}We demonstrate that several optomechanical systems can exhibit distinct spontaneous broken-symmetry phases, either continuous or discrete, yielding different coherent photon occupations in the ground state. By interacting with the two-level atoms, we find that the closing spectrum gap of the hybrid quantum system is determined by two distinct types of coupling degrees of freedom, which results in the emergence of a hybrid critical point. 

In addition, the hybrid critical point may display anomalous behavior when a drive with squeezed light is applied to the cavity. With appropriate squeezed parameters, it is possible to find two superradiant phases separated by this critical point. Considering that these two phases are independently induced by the two subsystems, respectively, they should be characterized by the two corresponding thermodynamic limits ($\eta$ and $N_a$), yielding two distinct order parameters. In this scenario, exploring whether such a system can undergo a second-order QPT between the two ordered phases with different broken symmetries would be an interesting topic beyond the Landau-Ginzburg-Wilson paradigm~\cite{2004deconfinedsenthil,2004Senthil,2017WangDeconfined}. With the impressive ongoing advancements in technology that allow for the realization of strong and even ultra-strong coupling between the cavity and a movable mirror, cavity optomechanical systems will become a potential platform for investigating critical phenomena.

\emph{Acknowledgments.}---We thank S. Yin and S. Ashhab for critical reading and stimulating discussions. This work was supported by the National Natural Science Foundation of China (Grant No. 12375025 and No. 11874432), the National Key R\&D Program of China (Grant No. 2019YFA0308200), and the China Postdoctoral Science Foundation (Grant No. 2021M693682). F. N. is supported in part by Nippon Telegraph and Telephone Corporation (NTT) Research, the Japan Science and Technology Agency (JST) [via the Quantum Leap Flagship Program (Q-LEAP) and the Moonshot R\&D Grant No. JPMJMS2061], the Asian Office of Aerospace Research and Development (AOARD) (via Grant No. FA2386-20-1-4069), and the Office of Naval Research (ONR).
	

\bibliography{QPTref.bib}

\begin{thebibliography}{77}
\bibitem [{ann({\natexlab{a}})}]{annotation2}%
  \BibitemOpen
  \href@noop {} {}\bibinfo {note} {We consider the resonant relationship
  between the cavity frequency and the atomic transition frequency. Moreover,
  using Hamiltonian $H_{\text{op}}$ in Eq.~\eqref{eq14} instead of $H$ to
  describe the optomechanical system can also obtain the same
  conclusions.}\BibitemShut {Stop}%
\end{thebibliography}

\onecolumngrid
\pagebreak
\widetext
\begin{center}
	\textbf{ {\large - Supplemental Material -} \\
		{\large Quantum Phase Transitions in Optomechanical Systems}}
\end{center}

\setcounter{equation}{0}
\setcounter{figure}{0}
\setcounter{table}{0}

\renewcommand{\theequation}{S\arabic{equation}}
\renewcommand{\thefigure}{S\arabic{figure}}
\renewcommand{\bibnumfmt}[1]{[S#1]}
\renewcommand{\citenumfont}[1]{S#1}

\section{\quad Photon occupations in ground state}
In this section, we interpret why the Hamiltonian,
\begin{equation}\label{Seq5}
	\widetilde{H}_{\text{om}}=a^{\dag}a-\xi^2 a^{\dag}aa^{\dag}a
\end{equation}
with eigenvalues
\begin{equation}\label{Seq3}
	\widetilde{E}_{\text{om}}=n-\xi^2 n^2,
\end{equation} 
does not have a well-defined ground state in the whole parameter space if the free term $a^{\dag}a$ is not considered as a perturbation. 

For a general Hamiltonian $H=H_0+\lambda H_{\text{I}}$, the interaction term $H_{\text{I}}$ would be conventionally considered as a perturbation if the coupling strength $\lambda$ is much smaller than the work frequency of the system. In this scenario, the ground state is mainly determined by the free term $H_0$. When increasing $\lambda$, $H_{\text{I}}$ becomes dominant, leading to the ground state being determined by both the free and interaction terms. 

The above description can be applied to the Dicke model and Rabi model. However, it is not true in Eq.~\eqref{Seq5} even though the parameter $\xi$ is extremely small. This is because we can always find an eigenstate $|n\rangle$ in $\widetilde{H}_{\text{om}}$ where its corresponding eigenvalue is smaller than that of the vacuum state $|0\rangle$ (which is, in fact, the lowest energy state of $a^{\dag}a$ ), regardless of the value of this parameter $\xi$. 

\begin{figure}[tpb]
	\centering
	\includegraphics[width = 0.5  \linewidth]{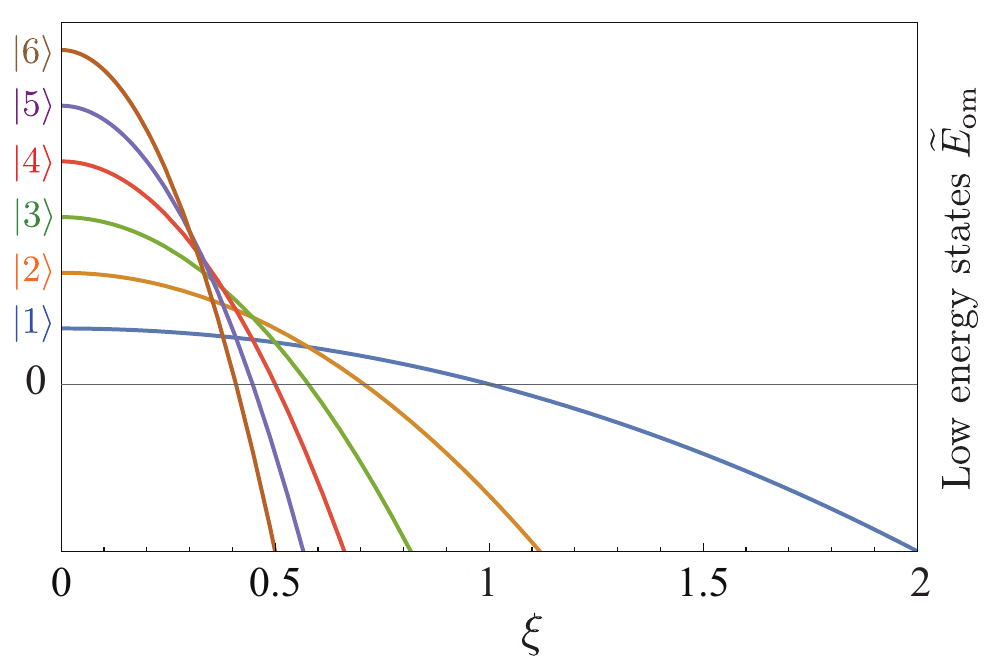}
	\caption{Level crossings of the low-energy states $\widetilde{E}_{\text{om}}$, according to Eq.~\eqref{Seq3}. As $n$ increases, the value of the intersection between the states $|n\rangle$ and $|0\rangle$ tends to zero, meaning that there is no well-defined ground state if the term $a^{\dag}a a^{\dag}a$ is considered as a perturbation.}
	\label{Vom_energylevel}	 	
\end{figure}

To clearly illustrate this point, we can evaluate the extrema of the energy,  
\begin{equation}\label{Seq6}
	\frac{d \widetilde{E}_{\text{om}}}{d\xi}=-2n^2 \xi=0~~~\quad\text{and}\quad~~~\frac{d^2 \widetilde{E}_{\text{om}}}{d \xi^2}=-2n^2<0,
\end{equation}
and find that for each value of $n$, the energy spectrum Eq.~\eqref{Seq3} has only a maximum value but no minimum value, as shown in Fig.~\ref{Vom_energylevel}. Additionally, Fig.~\ref{Vom_energylevel} shows that, as $n$ increases, the value of the intersection $\xi_0$ between $\widetilde{E}_{\text{om}}(n)$ and $\widetilde{E}_{\text{om}}=0$ (vacuum energy) will tend to zero. More precisely, by evaluating the formula $n-\xi_0^2 n^2=0$, it is not hard to find that when $n \rightarrow \infty, \xi_0 \rightarrow 0$. These analyses illustrate that no well-defined ground state can be determined by the term $a^{\dag}a$. 

If the free term $a^{\dag}a$ is considered as a perturbation, we can examine whether the ground state is determined by the quartic term $a^{\dag}aa^{\dag}a$. Under such a strategy, the negative half-axis of $\widetilde{E}_{\text{om}}$ in Fig.~\ref{Vom_energylevel} constitutes the component of the excitation spectrum of the system. Therefore, for $\xi \in [1,\infty]$, the lowest energy of system should be $\widetilde{E}_G=0$, giving the lowest level $|0\rangle$.

When $\xi$ decreases until $\xi=1$, there occurs a level crossing between the states $|0\rangle$ and $|1\rangle$, indicating that the perturbation term $a^{\dag}a$ gradually becomes dominant. For $\xi<1$, the lowest-energy state will meet a series of level crossings between the states $|n\rangle$ and $|n+1\rangle$, showing the instability. Finally, we obtain a well-defined ground state.

In the main text, the parameter $\xi$ becomes the dimensionless coupling strength $\kappa=\sqrt{\omega_m \omega_c}/g$. Due to the excitation spectrum on the negative half-axis of $\widetilde{E}_{\text{om}}$ shown in Fig.~\ref{Vom_energylevel}, performing the transformation $\widetilde{E}_{\text{om}} \rightarrow -\widetilde{E}_{\text{om}}$, which only changes the reference frame, can conveniently capture the physics, as shown in Figs.1(a), 2(a-c) in the main text.

\section{\quad Squeezed vacuum in the photon and phonon modes}\label{squeezed}

In this section, we show the squeezed vacuum of the cavity and mechanical modes at critical points, both of which have infinitely squeezed vacuum when $\eta \rightarrow \infty$ and have inseparable relation between the two squeezed states when $\eta$ is finite. First, by applying a displacement transformation with  $U_1=\exp[-(g/\omega_m)(b^{\dag}-b)]$ to the system Hamiltonian,
\begin{equation}\label{Seq9}
	H=\omega_c a^{\dag}a+\omega_m b^{\dag}b+g(a+a^{\dag})^{2}(b+b^{\dag}),
\end{equation}
the transformed Hamiltonian becomes
\begin{equation}\label{Seq10}
		\bar{H}= \omega_c a^{\dag}a+\omega_m b^{\dag}b-\frac{2g^2}{\omega_m}(a+a^{\dag})^2+\frac{g^2}{\omega_m } +g (a^2+a^{\dag 2})(b+b^{\dag})+2g a^{\dag}a(b+b^{\dag}).
\end{equation}
For convenience, this transformed Hamiltonian can also be rewritten as 
\begin{equation}\label{Seq11}
	\bar{H}_f= a^{\dag}a+\eta^{-1} b^{\dag}b-\frac{1}{4}\gamma^2(a+a^{\dag})^2+\frac{1}{8 }\gamma^2 +\frac{1}{2\sqrt{2} }\gamma \eta^{-\frac{1}{2}} (a^2+a^{\dag 2})(b+b^{\dag})+\frac{1}{\sqrt{2}  }\gamma \eta^{-\frac{1}{2}} a^{\dag}a(b+b^{\dag}),
\end{equation}
where $\gamma=2\sqrt{2}g/\sqrt{\omega_c \omega_m}$ is the dimensionless coupling strength and $\eta=\omega_c/\omega_m$. Note that the last two terms in Eq.~\eqref{Seq11} have a factor with a negative power of $\eta$. In the limit $\eta \rightarrow \infty$, the coefficients of these nonquadratic terms Eq.~\eqref{Seq11} become zero, leading to
\begin{equation}\label{Seq12}
	\bar{H}_f= a^{\dag}a-\frac{1}{4}\gamma^2(a+a^{\dag})^2+\frac{1}{8 }\gamma^2.
\end{equation}
Obviously, the Hamiltonian satisfies a $Z_2$ parity symmetry, which is obtained by eliminating the non-symmetry part with the classical limit. Further, equation \eqref{Seq12} can be diagonalized as
\begin{equation}\label{Seq13}
 \bar{H}_f=2\varepsilon_{\text{np}} d^{\dag}d+\frac{1}{8}\gamma^2+\varepsilon_{\text{np}}-\frac{1}{2}, 
\end{equation}
with $\varepsilon_{\text{np}}=\sqrt{(1/4)(1-\gamma^2)}$, which is vaild for $\gamma<1$ and vanishes at $\gamma=1$. The eigenstates of $\bar{H}_f$ for $\gamma <1$ are $|\psi \rangle(\gamma)=S(\xi(\gamma))|n\rangle|k\rangle$, where $S(x)=\exp[(x/2)(a^{\dag 2}-a^2)]$ and $\xi(\gamma)=-(1/4)\ln(1-\gamma^2)$.

From Eq.~\eqref{Seq13}, we can find an infinitely squeezed vacuum of cavity mode at $\gamma=1$, which implies an infinitely squeezed photon condensate in the ground state. It is not hard to imagine that the squeezing phenomenon of the cavity mode will give feedback to the mechanical mode by the radiation pressure. 

To further explore the underlying physics, one can employ a transformation with the unitary operator, $U_2=\exp[-(g/2\omega_c)(b+b^{\dag})(a^{\dag 2}-a^2)]$, to the Hamiltonian $\bar{H}$ in Eq.~\eqref{Seq10}.  The transformed Hamiltonian can be expressed as $\widetilde{H}=\widetilde{H}_e+\widetilde{H}_o$, where
\begin{equation}\label{Seq14}
\begin{split}
		\widetilde{H}_e= \,& \, \omega_c a^{\dag}a+\omega_m b^{\dag}b-\frac{1}{2!}\frac{g^2\omega_m}{2\omega_c^2}(a^{\dag 2}-a^2)^2+\sum_{n=1}^{\infty}\left[ \frac{1}{(2n)!}-\frac{1}{(2n-1)!}\right]2^{2n-3}\frac{g^{2n}}{\omega_c^{2n-1}}(b+b^{\dag})^{2n}(8a^{\dag}a+4)\\
		\,& \, -\sum_{n=0}^{\infty}\frac{1}{(2n)!}2^{2n+1}\frac{g^{2n+2}}{\omega_m \omega_c^{2n}}(b+b^{\dag})^{2n}(a^{\dag}+a)^2 -\sum_{n=1}^{\infty}\frac{1}{(2n-1)!}2^{2n-1}\frac{g^{2n}}{\omega_c^{2n-1}}(b+b^{\dag})^{2n}(a^{\dag 2}+a^2)+\frac{g^2}{\omega_m},\\
\end{split}
\end{equation}
which consists of even-order operators, while $\widetilde{H}_o$ consists of odd-order operators. 

This Hamiltonian $\widetilde{H}$ still cannot be diagonalized, but a variational method can be used for analysis. We propose a trial wave function $|\psi(r,s)\rangle=\mathcal{S}_a(r)\mathcal{S}_b(s)|0_a,0_b\rangle$ for $\widetilde{H}$, where $\mathcal{S}_y(x)=\exp[(x^2/2)(y^{\dag 2}-y^2)]$ is the squeezing operator with a bosonic operator $y\in[a,b]$ and a variational parameter $x \in [r,s]$. The energy function can be obtained as follows
\begin{equation}\label{Seq15}
	\begin{split}
		\widetilde{E}(r,s)= \,& \, \sinh^2(r)+\eta^{-1}\sinh^2(s)+\sum_{n=1}^{\infty}\left[ \frac{1}{(2n)!}-\frac{1}{(2n-1)!}\right]2^{2n-3}\frac{1}{8^n}\gamma^{2n}\eta^{-n}\frac{(2n)!}{2^n n!}e^{2ns}[8\sinh^2(r)+4]+\frac{1}{8}\gamma^2\\
		\,& \, -\sum_{n=0}^{\infty}\frac{1}{(2n)!}2^{2n+1}\frac{1}{8^{n+1}}\gamma^{2n+2}\eta^{-n}\frac{(2n)!}{2^n n!}e^{2ns}e^{2r}-\sum_{n=1}^{\infty} \frac{1}{(2n-1)!}2^{2n-1}\frac{1}{8^n}\gamma^{2n}\eta^{-n}\frac{(2n)!}{2^n n!}e^{2ns}\sinh(2r),
	\end{split}
\end{equation}
where $\widetilde{E}(r,s)$ has been renormalized by $\omega_c$. By minimizing the energy function with respect to the $r$ and $s$, we can obtain
\begin{equation}\label{Seq16}
	\exp(4r)=\frac{1+A_1+C_1}{1+A_1-B_1-C_1},
\end{equation}
where 
\begin{equation}\label{Seq16_1}
	\begin{split}
	\,& \, A_1=\sum_{n=1}^{\infty}\left[ \frac{1}{(2n)!}-\frac{1}{(2n-1)!}\right]2^{2n-3}\frac{1}{8^{n-1}}\gamma^{2n}\eta^{-n}\frac{(2n)!}{2^n n!}e^{2ns},\\
	\,& \, B_1=\sum_{n=0}^{\infty}\frac{1}{(2n)!}2^{2n+3}\frac{1}{8^{n+1}}\gamma^{2n+2}\eta^{-n}\frac{(2n)!}{2^n n!}e^{2ns},\\
	\,& \, C_1=\sum_{n=1}^{\infty} \frac{1}{(2n-1)!}2^{2n}\frac{1}{8^n}\gamma^{2n}\eta^{-n}\frac{(2n)!}{2^n n!}e^{2ns},
	\end{split}
\end{equation}
and 
\begin{equation}\label{Seq17}
	\exp(4s)=\frac{1+A_2}{1-\gamma^2\left[\sinh^2(r)+\frac{1}{2}+\sinh(2r)+\frac{1}{4}\gamma^2e^{2r}\right]},
\end{equation}
where 
\begin{equation}\label{Seq17_1}
	\begin{split}
		A_2=\,& \, -\left[8\sinh^2(r)+4\right]\sum_{n=2}^{\infty}\left[ \frac{1}{(2n)!}-\frac{1}{(2n-1)!}\right]2^{2n-2}\frac{1}{8^{n}}\gamma^{2n}\eta^{1-n}\frac{(2n)!}{2^n n!}e^{2ns+2s}\\
		\,& \, +e^{2r}\sum_{n=2}^{\infty}\frac{1}{(2n)!}2^{2n+2}\frac{1}{8^{n+1}}\gamma^{2n+2}\eta^{1-n}\frac{(2n)!}{2^n n!}e^{2ns+2s}2n\\
		\,& \, +\sinh(2r)\sum_{n=2}^{\infty} \frac{1}{(2n-1)!}2^{2n}\frac{1}{8^n}\gamma^{2n}\eta^{1-n}\frac{(2n)!}{2^n n!}e^{2ns+2s}2n.
	\end{split}
\end{equation}
Now we have derived the results of the squeezed vacuum of photon and phonon mode, as shown in Eqs.~\eqref{Seq16} and~\eqref{Seq17}.

When $\eta \rightarrow \infty$, Eqs.~\eqref{Seq16} and~\eqref{Seq17} can be rewritten as 
\begin{equation}\label{Seq18_1}
		\exp(4r)= \frac{1}{1-\gamma^2}
\end{equation}
and
\begin{equation}\label{Seq18_2}
	\exp(4s)= \frac{1}{1-\gamma^2\left[\sinh^2(r)+\frac{1}{2}+\sinh(2r)+\frac{1}{4}\gamma^2e^{2r}\right]},
\end{equation}
which are given in the main text. Note that after implementing the unitary transformation $U^{\dag}_2\bar{H}U_2$, we obtain not only the result in Eq.~\eqref{Seq18_1}, which is consistent with that in Eq.~\eqref{Seq13}, but also the squeezing information of the mechanical mode shown in Eq.~\eqref{Seq18_2}, which is induced by the squeezed field of the cavity mode. 

It is clear that upon displacing the mechanical mode by $g/\omega_m$, the squeezed vacuum of cavity mode could be induced and subsequently gives feedback to the mechanical mode via radiation pressure. Notably, such feedback would not once again transfer to the cavity mode in the classical limit $\eta \rightarrow \infty$, Eq.~\eqref{Seq18_1}, except for finite $\eta$, Eq.~\eqref{Seq16}.

In the main text, without loss of generality, we have chosen the energy function $\widetilde{E}(r,s)$ up to the fourth order for analyzing the case of $\eta \rightarrow \infty$ and finite $\eta$, whose conclusions are the same as that of Eqs.~\eqref{Seq16} and~\eqref{Seq17}.


\section{\quad Modifying a QPT in optomechanical system }

In the previous section, Eq.~\eqref{Seq13} showed that the normal phase occurs for $\gamma < 1$. In this section, we will show that when the system is driven by a squeezed field of the cavity mode with a suitable squeezing parameter, the occurrence of the normal phase can be allowed to occur in the region $\gamma >1$, and the coupling strength required to reach the critical point can be significantly reduced.

By employing a transformation with the unitary operator, $S_{\zeta}=\exp{[(1/2)(\zeta^{*}a^2-\zeta a^{\dag 2})]}$, where $\zeta=\xi e^{i\theta}$ represents the squeezed parameter, to the system Hamiltonian $H$ in Eq.~\eqref{Seq9}, with the squeezing driven term $\xi(a^{\dag 2}e^{-i\theta}+a^2 e^{i\theta})$, we can obtain a transformed Hamiltonian 
\begin{equation}\label{Seq19_0}
\begin{split}
	H_{(\theta,\xi)}=\,& \, \frac{\omega_c}{2}\cosh(2\xi)(2a^{\dag}a+1)-\frac{\omega_c}{2}\sinh(2\xi)(e^{i\theta}a^{\dag 2}+e^{-i\theta}a^2)-\frac{\omega_c}{2}\\
 \,& \,+\omega_m b^{\dag}b+g (b+b^{\dag})\left[\cosh{(\xi)}(a+a^{\dag})-\sinh{(\xi)}(e^{-i\theta}a+e^{i\theta}a^{\dag})\right]^2\\
 \,&\,+S^{\dag}_{\zeta}\left[\xi(a^{\dag 2}e^{-i\theta}+a^2e^{i\theta})\right]S_{\zeta},
 \end{split}
\end{equation}
which depends on the squeezing direction $\theta$ and the squeezing amplitude $\xi$. 

When the squeezing direction is $\theta=0$, the transformed Hamiltonian can be written as 
\begin{equation}\label{Seq19}
\begin{split}
	H_{(\theta=0,\xi)}=\,&\, \frac{\omega_c}{2}\cosh(2\xi)(2a^{\dag}a+1)-\frac{\omega_c}{2}\sinh(2\xi)(a^{\dag 2}+a^2)-\frac{\omega_c}{2}\\
 \,&\,+\omega_m b^{\dag}b+g e^{-2\xi}(a+a^{\dag})^2(b+b^{\dag})+S^{\dag}_{(\theta=0,\xi)}\left[\xi(a^{\dag 2}+a^2)\right]S_{(\theta=0,\xi)}.
 \end{split}
\end{equation}
In order to find the anti-squeezing term of the cavity mode, which can induce a singularity of the intrinsic squeezed vacuum, one can perform a displacement transformation with $\widetilde{U}_{\theta \rightarrow 0}=\exp[-(g e^{-2\xi}/\omega_m)(b^{\dag}-b)]$, to the Hamiltonian $H_{(\theta=0,\xi)}$. The transformed Hamiltonian becomes
\begin{equation}\label{Seq20}
	\begin{split}
		\widetilde{H}_{(\theta=0,\xi)}=\,& \, \frac{1}{2}\cosh(2\xi)(2a^{\dag}a+1)-\frac{1}{2}\sinh(2\xi)(a^{\dag 2}+a^2)-\frac{1}{2}+\eta^{-1} b^{\dag}b-\frac{1}{4}\gamma^2 e^{-4\xi}(a+a^{\dag})^2\\
		\,& \, +\frac{1}{2\sqrt{2}}\gamma \eta^{-\frac{1}{2}} e^{-2\xi}(a^2+a^{\dag 2})(b+b^{\dag})+\frac{1}{\sqrt{2}}\gamma \eta^{-\frac{1}{2}}e^{-2\xi}a^{\dag}a(b+b^{\dag})+\frac{1}{8}\gamma^2 e^{-4\xi}\\
            \,&\, +\eta^{-1}\frac{1}{\omega_m}S^{\dag}_{(\theta=0,\xi)}\left[\xi(a^{\dag 2}+a^2)\right]S_{(\theta=0,\xi)},
	\end{split}
\end{equation}
where $\widetilde{H}_{(\theta=0,\xi)}$ has been renormalized by the cavity frequency $\omega_c$. In the classical limit, $\eta \rightarrow \infty$, Eq.~\eqref{Seq20} can be written as
\begin{equation}\label{Seq21}
	\begin{split}
		\widetilde{H}_{(\theta=0,\xi)}=\,& \, \frac{1}{2}\cosh(2\xi)(2a^{\dag}a+1)-\frac{1}{2}\sinh(2\xi)(a^{\dag 2}+a^2)-\frac{1}{4}\gamma^2 e^{-4\xi}(a+a^{\dag})^2+\frac{1}{8}\gamma^2 e^{-4\xi}-\frac{1}{2}.
	\end{split}
\end{equation}
Equation~\eqref{Seq21} can be diagonalized giving
\begin{equation}\label{Seq22}
		\widetilde{H}_{(\theta=0,\xi)}=2\varepsilon_{\xi} d^{\dag}d+\frac{1}{8}\gamma^2 e^{-4\xi}+\varepsilon_{\xi} -\frac{1}{2},
\end{equation}
with 
\begin{equation}\label{Seq23}
	\varepsilon_{\xi}=\frac{1}{2}\sqrt{1-\gamma^2 \exp(-2\xi)}.
\end{equation}
When $\xi = 0$, the result of Eq.~\eqref{Seq23} is the same as that of Eq.~\eqref{Seq13}. However, for $\xi =2\ln(\gamma)$, Eq.~\eqref{Seq23} becomes 
\begin{equation}\label{Seq24}
	\varepsilon_{\xi \rightarrow 2\ln(\gamma)}=\frac{1}{2}\sqrt{1-\gamma^{-2}},
\end{equation}
which is real only for $\gamma \geq 1$ and vanishes at $\gamma=1$. This result shows that exploiting a squeezed field of the cavity mode with an appropriate squeezing parameter $\zeta$ to drive the system can alter the region where the normal phase occurs. Moreover, by employing the strategy used in the section \textquotedblleft superradiant phase\textquotedblright of the main text to Eq.~\eqref{Seq19}, we can also determine that the corresponding superradiant phase occurs in the region $\gamma < 1$ for the same squeezing amplitude $\xi =2\ln(\gamma)$.

Now we discuss the case when the squeezing direction is chosen as $\theta=\pi$. The transformed Hamiltonian becomes
\begin{equation}\label{Seq24.5}
\begin{split}
	H_{\theta=\pi,\xi}=\,&\, \frac{\omega_c}{2}\cosh(2\xi)(2a^{\dag}a+1)+\frac{\omega_c}{2}\sinh(2\xi)(a^{\dag 2}+a^2)-\frac{\omega_c}{2}\\
          \,&\, +\omega_m b^{\dag}b+g e^{2\xi}(a+a^{\dag})^2(b+b^{\dag})+S^{\dag}_{(\theta=\pi,\xi)}\left[\xi(a^{\dag 2}+a^2)\right]S_{(\theta=\pi,\xi)}.
\end{split}
\end{equation}
After implementing the above similarity transformation, we can obtain
\begin{equation}\label{Seq24.6}
		\widetilde{H}_{(\theta=\pi,\xi)}=2\varepsilon_{(\theta=\pi,\xi)} d^{\dag}d+\frac{1}{8}\gamma^2 e^{4\xi}+\varepsilon_{(\theta=\pi,\xi)} -\frac{1}{2},
\end{equation}
with 
\begin{equation}\label{Seq24.7}
	\varepsilon_{(\theta=\pi,\xi)}=\frac{1}{2}\sqrt{1-\gamma^2 \exp(2\xi)}.
\end{equation}
This result shows that increasing the squeezing amplitude $\xi$ can exponentially reduce the coupling strength $g$ required to reach the critical point.

In addition, these results do not apply to the Rabi or Dicke models, as the cavity fields of these two models have no deterministic symmetry. Therefore, if one controllable squeezed field of the cavity mode is exploited to drive the hybrid quantum system described in the main text, it is possible to find two superradiant phases, which are induced by the optomechanical system and light-atom system, respectively, and separated by the hybrid critical point. The two ordered phases could be characterized by two types of thermodynamic limits. Therefore, they belong to distinct symmetry-broken phases.  In this scenario, it naturally arises the exciting topic of whether the hybrid system can undergo a direct second-order QPT between the two ordered phases beyond the Landau-Ginzburg-Wilson paradigm.



\section{\quad Quantum phase transitions in hybrid quantum systems}

In this section, we will examine the characteristics of the superradiant phase in the hybrid quantum system. To determine the superradiant phase, we must find the macroscopic coherence of the cavity mode in the ground state. Generally, when the ground-state energy has been weighted by one thermodynamic limit, the macroscopic coherence of the cavity mode can be evaluated using the mean-field approach.  In the case of a hybrid quantum system, the macroscopic coherence would be determined by the two types of thermodynamic limits, where one of them is described by $\omega_c/\omega_m \rightarrow \infty$ and the other is described by $N \rightarrow \infty$. 

Due to the possibility that these two limits may affect the nontrivial phase of the system, either competitively or independently, finding the solutions of macroscopic coherence in the ground-state energy using the mean-field approach with the two limits is very complicated. However, suppose we need to roughly capture the potential characteristics of the superradiant phase in the hybrid quantum system. In this case, we may only consider one of the two limits in the mean-field energy, while the other could become a large constant. Here, we will only consider the limit, $N \rightarrow \infty$, for the mean-field approach.

The hybrid quantum system can be described by
\begin{equation}\label{Seq25}
	H_{\text{h}} =\omega_c a^{\dag}a+\omega_m b^{\dag}b+g(a+a^{\dag})^{2}(b+b^{\dag}) +\omega_a J_z+\frac{\lambda}{\sqrt{N_a}}(a+a^{\dag})(J_{+}+J_{-})+\frac{\alpha \lambda^2}{\omega_a}(a+a^{\dag})^2,
\end{equation}
where the cavity frequency and the atomic transition frequency are in resonance, namely, $\omega_c = \omega_a$. Note that using the Hamiltonian $H_{\text{op}}$ in Eq.(10) in the main text, instead of $H=\omega_c a^{\dag}a+\omega_m b^{\dag}b+g(a+a^{\dag})^{2}(b+b^{\dag})$ to describe the optomechanical system, we can also reach the same conclusions. By displacing the mechanical mode with the single-photon coupling strength $g/\omega_m$, the system Hamiltonian can be transformed to
\begin{equation}\label{Seq26}
	H_{\text{h}} =\omega_c a^{\dag}a+\omega_m b^{\dag}b+g(a^2+a^{\dag 2}+2a^{\dag 2}a)(b+b^{\dag})+\left(\frac{\alpha \lambda^2}{\omega_a}-\frac{2g^2}{\omega_m}\right)(a+a^{\dag})^2+\frac{g^2}{\omega_m} +\omega_a J_z+\frac{\lambda}{\sqrt{N_a}}(a+a^{\dag})(J_{+}+J_{-}).
\end{equation}
After the displacement, the cavity field can reach a singularity induced by the two terms, $-2g^2/\omega_m(a+a^{\dag})^2$ and $(\lambda/\sqrt{N_a})(a+a^{\dag})(J_{+}+J_{-})$, together. However, the latter will be suppressed if we consider the $\bm A^2$ term. Now, we first consider the limit, $\omega_c/\omega_m \rightarrow \infty$, Eq.~\eqref{Seq26} can be written as  
\begin{equation}\label{Seq27}
	\widetilde{H}_{\text{h}} =a^{\dag}a+\left(\frac{\alpha \lambda^2}{\omega_a \omega_c}-\frac{2g^2}{\omega_m \omega_c}\right)(a+a^{\dag})^2+\frac{g^2}{\omega_m \omega_c } +\frac{\omega_a}{\omega_c } J_z+\frac{\lambda}{\omega_c \sqrt{N_a}}(a+a^{\dag})(J_{+}+J_{-}),
\end{equation}
where $\widetilde{H}_{\text{h}} $ is renormalized by $\omega_c$. 

By applying a squeezing transformation, $a=\cosh(r)d+\sinh(r)d^{\dag}$ with a squeezing parameter $r=(-1/4)\ln(1+\alpha\mu^2-\gamma^2)$ ($\mu=2\lambda/\sqrt{\omega_a \omega_c}$ is a dimensionless coupling strength of light-atoms interacting system), the Hamiltonian $\widetilde{H}_{\text{h}}$ can be written in a more compact form, 
\begin{equation}\label{Seq28}
	\widetilde{H}_{\text{h}} =\widetilde{\omega}_c d^{\dag} d +\omega_a J_z+\frac{\widetilde{\lambda}}{\sqrt{N_a}}(d+d^{\dag})(J_{+}+J_{-})+\widetilde{Q},
\end{equation}
where $\widetilde{\omega}_c=\omega_c e^{-2r}$, $\widetilde{\lambda}=\lambda e^{r}$, and $\widetilde{Q}=(\omega_c/2)(e^{-2r}-1)+g^2/\omega_m$. By using the Holstein-Primakoff approach with the transformation $J_{+}=d^{\dag}\sqrt{N_a-d^{\dag}d}$, $J_{-}=\sqrt{N_a-d^{\dag}d}d$, and $J_z=d^{\dag}d-N_a/2$, Eq.~\eqref{Seq28} becomes
\begin{equation}\label{Seq29}
	\widetilde{H}_{\text{h(np)}} = \widetilde{\omega}_c d^{\dag} d+\omega_a (c^{\dag}c-\frac{N_a}{2})+\frac{\widetilde{\lambda}}{\sqrt{N_a}}\left(d+d^{\dag}\right)\left(c^{\dag} \sqrt{N_a-c^{\dag}c}+\sqrt{N_a-c^{\dag}c}~c \right)+\tilde{Q},
\end{equation}
whose energy reads
\begin{equation}\label{Seq30}
	\epsilon_{\pm}^{\text{np}}=\sqrt{\frac{1}{2}\left(\widetilde{\omega}_c^2+\omega_a^2\pm\sqrt{(\widetilde{\omega}_c^2-\omega_a^2)^2+16\widetilde{\lambda}^2\widetilde{\omega}_c\omega_a}\right)}.
\end{equation}
The excitation energy of the lowest branch $\varepsilon_{-}$ vanishes at $\mu^2(1-\alpha)+\gamma^2=1$, locating the quantum critical point for the hybrid quantum system, as shown in Eq.(15) in the main text. 

In the following, we can determine the superradiant phase of the system by only considering the thermodynamic limit, $N \rightarrow \infty$, in the mean-field approach. Through displacing the bosonic operator with respect to their mean value, i.e., $d \rightarrow d+\sqrt{N_a}\zeta$ and $c \rightarrow c+\sqrt{N_a}\beta$, for the Hamiltonian $\widetilde{H}_{\text{h(np)}}$, we can derive the Hamiltonian $\widetilde{H}_{\text{h}(\text{sp})}$, whose ground state energy reads
\begin{equation}\label{Seq31}
	E_G=N_a \omega_a|\beta|^2+N_a \widetilde{\omega}_c|\zeta|^2+4 \widetilde{\lambda} N_a \zeta \beta \sqrt{1-\beta^2}-\frac{N_a}{2} \omega_a.
\end{equation}
By minimizing the ground state energy with respect to $\alpha$ and $\beta$, we can obtain
\begin{equation}\label{Seq32}
	\begin{split}
	\,& \, \beta=\pm\sqrt{\frac{1}{2}\left(1-\widetilde{\delta}^{-2}\right)}, \\
	\,& \, \zeta=\mp\sqrt{\frac{\omega_a}{4\widetilde{\omega}_c}}\sqrt{\widetilde{\delta}^2-\widetilde{\delta}^{-2}},
	\end{split}
\end{equation}
where $\widetilde{\delta}=2\widetilde{\lambda}/\sqrt{\omega_a \widetilde{\omega}_c}$ is a dimensionless coupling strength. The nonzero value of $\zeta$ can be found for $\widetilde{\delta}>1$ and indicates a nonzero coherence of the boson field in the ground state, which is an order parameter of the superradiant phase transition. According to those nonzero coherences in Eq.~\eqref{Seq32}, the Hamiltonian of the superradiant phase can be written as 
\begin{equation}\label{Seq33}
	\widetilde{H}_{\text{h(sp)}}= \widetilde{\omega}_c d^{\dag} d+\left(\omega_a-\frac{2\widetilde{\lambda}\zeta \beta}{\sqrt{1-\beta^2}}\right) c^{\dag}c+\left(\widetilde{\lambda}\sqrt{1-\beta^2}-\frac{\widetilde{\lambda}\beta^2}{\sqrt{1-\beta^2}}\right)(c+c^{\dag})(d+d^{\dag})-\left(\frac{\zeta\beta\widetilde{\lambda}}{\sqrt{1-\beta^2}}+\frac{\widetilde{\lambda}\zeta\beta^3}{2\left(1-\beta^2\right)^{\frac{3}{2}}}\right)(c+c^{\dag})^2,
\end{equation}
whose energy spectrum reads
\begin{equation}\label{Seq34}
	\epsilon_{\pm}^{\text{sp}}=\sqrt{\frac{1}{2}\left(\widetilde{\omega}_c^2+\widetilde{\delta}^4\omega_a^2\pm\sqrt{(\widetilde{\omega}_c^2-\widetilde{\delta}^4\omega_a^2)^2+4\widetilde{\omega}_c^2\omega_a^2}\right)}.
\end{equation}

\begin{figure}[htp]
	\centering
	\includegraphics[width = 0.8  \linewidth]{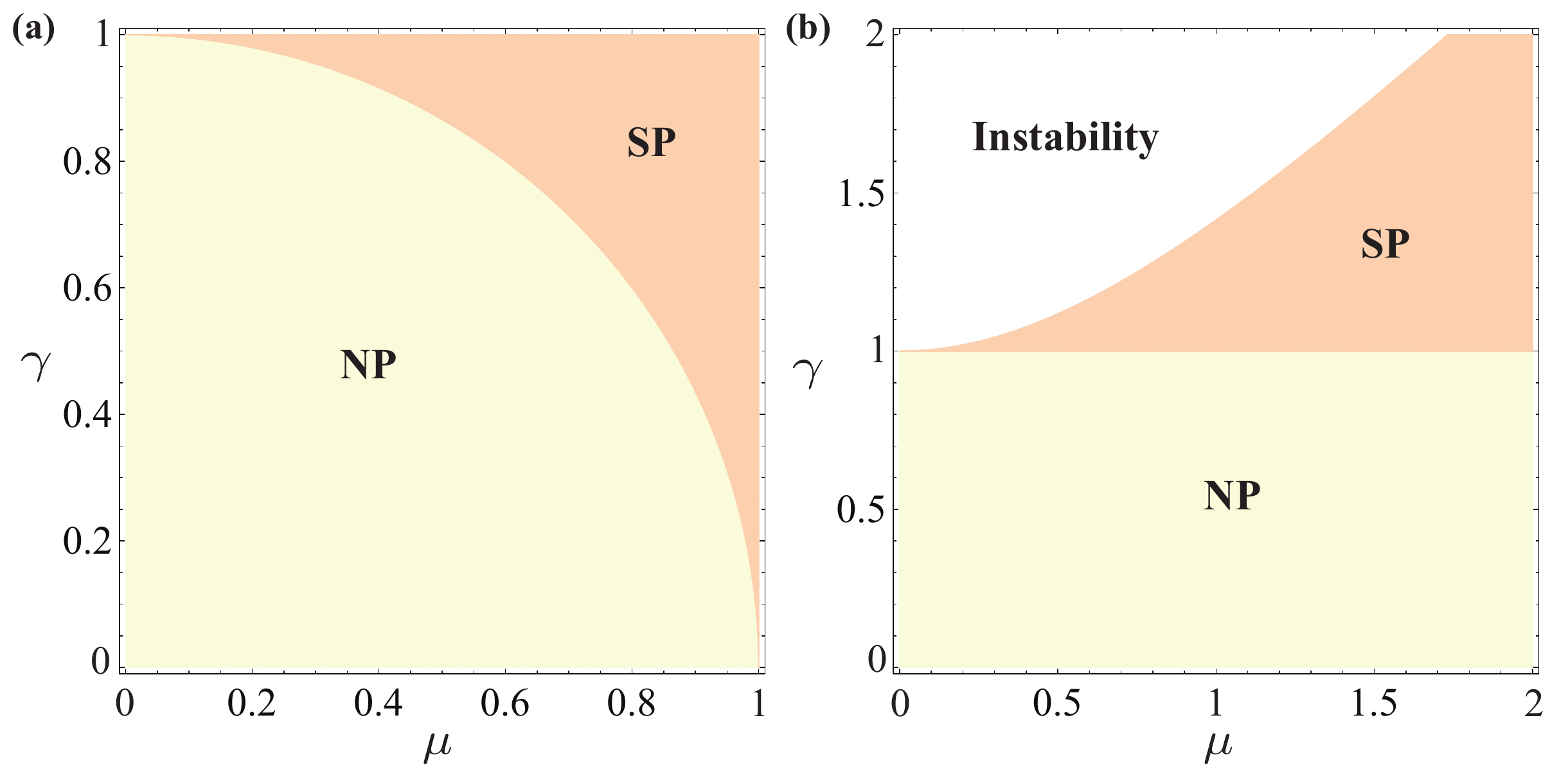}
	\caption{Phase diagram of the hybrid quantum system in the ($\mu$,$\gamma$) plane, described by Eq.~\eqref{Seq30} and Eq.~\eqref{Seq34}. (a) $\alpha=0$: the boundary (critical line) between the superradiant phase (SP) and the normal phase (NP) is characterized by $\mu^2+\gamma^2=1$. (b) $\alpha=1$: the boundary is given by $\gamma^2=1$.}
	\label{HybridQuantumSystem}	 	
\end{figure}

From the results of the mean-field in Eq.~\eqref{Seq32}, we can also determine that $\widetilde{\delta}=1$ is the critical point of the system, which can be expanded as
\begin{equation}\label{Seq35}
	\mu^2(1-\alpha)+\gamma^2=1.
\end{equation}
When $\alpha=0$ (meaning the absence of the $\bm A^2$ term), Eq.~\eqref{Seq35} becomes $\mu^2+\gamma^2=1$, indicating a \textit{hybrid} critical point. Such a point features the boundary between the normal and superradiant phases. Therefore, based on the energy spectrum [Eq.~\eqref{Seq30} and Eq.~\eqref{Seq34}] or the nonzero coherence $\zeta$, we can plot the phase diagram shown in Fig.~\ref{HybridQuantumSystem}(a), where the boundary has been characterized by the hybrid critical point $\mu^2+\gamma^2=1$. When $\alpha=1$ (meaning including the $\bm A^2$ term), the critical point of the system is $\gamma^2=1$ and the corresponding phase diagram is displayed in Fig.~\ref{HybridQuantumSystem}(b). In this case, we find that the region of the superradiant phase can still be affected by the light-atom interaction ($\mu$), even though the critical point is only dominated by the optomechanical system ($\gamma$).

\end{document}